\documentclass[useAMS]{mn2e}  
\usepackage{graphicx}

\title[Stratifications in magnetic Ap stars]
      {Empirical chemical stratifications in magnetic Ap stars: 
       questions of uniqueness}

\author[M.J.~Stift, G.~Alecian]
       {M.J.~Stift$^{1,2}$,
        G.~Alecian$^2$\\
        $^1$Institut f{\"u}r Astronomie (IfA), Universit{\"a}t Wien,
            T{\"u}rkenschanzstrasse 17, A-1180 Wien, Austria\\
        $^2$LUTH, Observatoire de Paris, CNRS, Universit{\'e} Paris
            Diderot, 5 place Jules Janssen, F-92190 Meudon, France}
\begin{document}

\date{Accepted 2008}

\pagerange{\pageref{firstpage}--\pageref{lastpage}} \pubyear{2009}

\maketitle

\label{firstpage}

\begin{abstract}
 {Over the last decades, modelling of the inhomogeneous vertical
  abundance distributions of various chemical elements in magnetic 
  peculiar A-type has largely relied on simple step-function 
  approximations. In contrast, the recently introduced regularised 
  vertical inverse problem (VIP) is not based on parametrised 
  stratification profiles and has been claimed to yield unique 
  solutions without a priori assumptions as to the profile shapes.
  It is the question of uniqueness of empirical stratifications which 
  is at the centre of this article. An error analysis establishes 
  confidence intervals about the abundance profiles and it is shown 
  that many different step-functions of sometimes widely different 
  amplitudes give fits to the observed spectra which equal the VIP 
  fits in quality. Theoretical arguments are advanced in favour of 
  abundance profiles that depend on magnetic latitude, even in 
  moderately strong magnetic fields. Including cloud, cap and ring 
  models in the discussion, it is shown that uniqueness of solutions 
  cannot be achieved without phase resolved high signal-to-noise 
  ratio (S/N) and high spectral resolution (R) spectropolarimetry 
  in all 4~Stokes parameters.
} 
   \end{abstract}


\section{Introduction}
\label{sec:intro}

Over the years, evidence has accumulated that a number of 
(magnetic) Ap stars not only exhibit non-uniform distributions 
of chemical elements over their surfaces -- first abundance maps 
within the framework of the oblique rotator model date back to 
Deutsch (1958) and to Pyper (1969) -- but that vertical abundance 
distributions in their atmospheres also are non-uniform. Such 
stratified abundances are to be expected from diffusion theory 
(Michaud 1970) and are reflected in spectra that cannot be fitted 
by constant elemental abundances. Depending on ionisation stage 
and on excitation potential, different abundances may be needed 
to fit different spectral lines of a given element. Borsenberger 
et al. (1981) were the first to compare (with some success) 
theoretical equivalent widths based on predicted stratified Ca and 
Sr abundances to observed equivalent widths, and Alecian (1982) 
attempted to detect Mn stratification in the atmosphere of 
$\nu$ Her. In a recent review paper, Ryabchikova (2008) has 
summarised recent results on abundance stratifications, illustrating 
her discussion with many empirical and some theoretical profiles 
for various chemical elements in different Ap stars. The vast
majority of these curves essentially correspond to step functions,
with lower abundances in the outer layers and a (sometimes drastic) 
increase towards the deeper layers -- in a few cases just the
opposite behaviour is found. These abundance jumps can range from 
a few 0.1\,dex to 4 dex and more; some inversion codes yield
smooth curves, whereas in others they are assumed to be more or less
discontinuous. It has, however, always been assumed that the profiles do 
not depend on the direction or on the strength of the local magnetic 
field, so that in a given star the same profile applies everywhere,
regardless of the variations in magnetic field direction and in field 
strength as, for example, found in a dipolar geometry.

On the other hand, Alecian \& Stift (2006) have presented snapshots of
abundance increases as a function of magnetic field direction which
reveal sometimes huge differences between vertical and horizontal
fields. Alecian \& Stift (2008) have also shown that, depending on 
the field direction, equilibrium stratifications can differ by several 
dex in the upper layers. This result is consistent with what has already 
been known about the sensitivity of the diffusion velocity to the 
horizontal component of the magnetic field (see e.g. Babel \& Michaud
1991ab), and it is certainly not at variance with the apparent 
correlations observed between abundance patches and magnetic geometries 
in magnetic oblique rotators (see e.g. Kochukhov et al. 2002).

So far, no attempts have been made to reconcile the empirical modelling
of stratifications with the sometimes complex abundance structures 
predicted for magnetic stellar atmospheres by theoretical studies. 
Kochukhov et al. (2006) (henceforth KTR06) rather have introduced a 
new empirical approach
based on a regularised solution of the vertical inversion problem (VIP).
Their abundance profiles have allegedly been derived ``without making a 
priori assumptions about the shape of chemical distributions'' and their 
``optimum regularisation'' is claimed to ensure the uniqueness of the 
solution. But is it really possible that this new and radically 
empirical approach yields the answers that theory cannot yet provide? 
Is the VIP method assumption-free or is it still 
subject to some hidden constraints?  Where in the atmosphere are the 
empirical stratifications well defined, and can they be truly considered
unique? How small are the details that a method based on high resolution 
Stokes\,$I$ spectra can reliably detect? Can alternative step-function 
like solutions be found for HD\,133792, perhaps even solutions that 
depend on magnetic latitude in an oblique rotator model? What is the 
kind of information that can reliably be gleaned from such inversions?

\begin{figure}
\includegraphics[width=84mm]{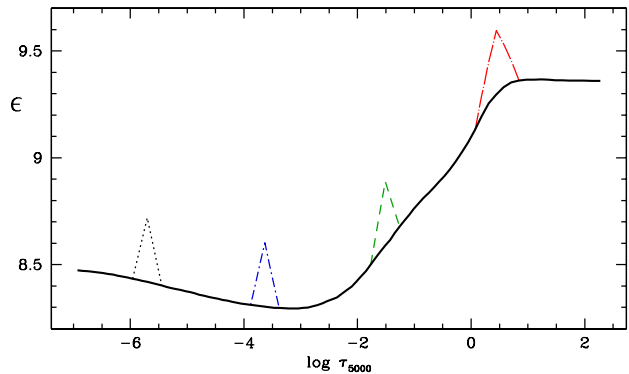}
\caption{The VIP based Fe stratification profile for HD\,133792 is 
plotted as logarithm of abundance $\epsilon$ (normalised to 
$\log H = 12.0$) versus $\log \tau_{5000}$. The selection of 
perturbations to this profile which underly the results presented 
in Fig.\,\ref{fig2} pertain to $\log \tau = -5.701$~(dot), 
-3.632~(dot - short dash), -1.511~(short dash), and 
+0.439~(dot - long dash).}
\label{fig1}
\end{figure}

This paper addresses these questions (and a few more). A simple but
realistic error analysis makes it possible to estimate the interval
in optical depth over which the empirical stratification profiles are 
more or less well defined. Extensive numerical modelling (involving
models based on cap-, ring- and cloud-like structures) is then used 
in the assessment of the question whether uniqueness of the abundance 
profiles can be attained with the methods and data presently at hand. 
Finally we advance ideas for a strategy that could remove some 
non-uniqueness of the models and lead to more reliable stratification 
results.

\section{Empirical stratifications}

From the plots presented by Ryabchikova (2008) in her review which 
are based on results for magnetic and non-magnetic Ap stars taken 
from recent literature, it emerges that practically all empirical 
profiles correspond to a step-function described by 4 parameters, 
viz. the abundance in the upper atmosphere, the abundance in the 
deep layers, the position of the jump and the width of the jump. 
These stratification profiles have always been assumed to remain
constant over the star, regardless of the strength of the stellar 
magnetic field. It is true that in stars with weak fields
profiles do not depend on magnetic latitude over large parts of 
the stellar atmosphere -- in a 1\,kG horizontal field Alecian 
\& Stift (2007) have found differences compared to the zero field 
case only for $\log \tau_{\rm 5000} < -3$ -- but globally constant 
profiles can certainly not be expected at 10 or 20\,kG as, for 
example, encountered in HD\,144897 and in HD\,66318. 
Notice that for elements having very low abundances, such as the 
rare earths in most stars, radiative accelerations generally 
exceed gravity by a considerable amount (this is due to the fact 
that absorption lines are completely unsaturated and remain 
unsaturated even for relatively strong element enhancements). 
In stable atmospheres, these elements then experience very high 
diffusion velocities and will be expelled from the star, except 
if they are blocked by a magnetic field. This generally occurs 
high up in the atmosphere ($\log \tau  < -3$) in places where 
magnetic field lines are horizontal (Alecian \& Stift, 2009, in 
preparation) and holds true even for weak magnetic fields.
Evidence for the existence of increased Nd abundances above 
$\log \tau = -3.5$ in $\gamma$\,Equ and in HD\,24712 presented
by Mashonkina, Ryabchikova \& Ryabtsev (2005) is certainly not
at variance with the theoretical prediction and so even for the 
1\,kG case one should expect significant horizontal differences 
in the vertical stratifications of some ions.

\begin{figure*}
\includegraphics[width=176mm]{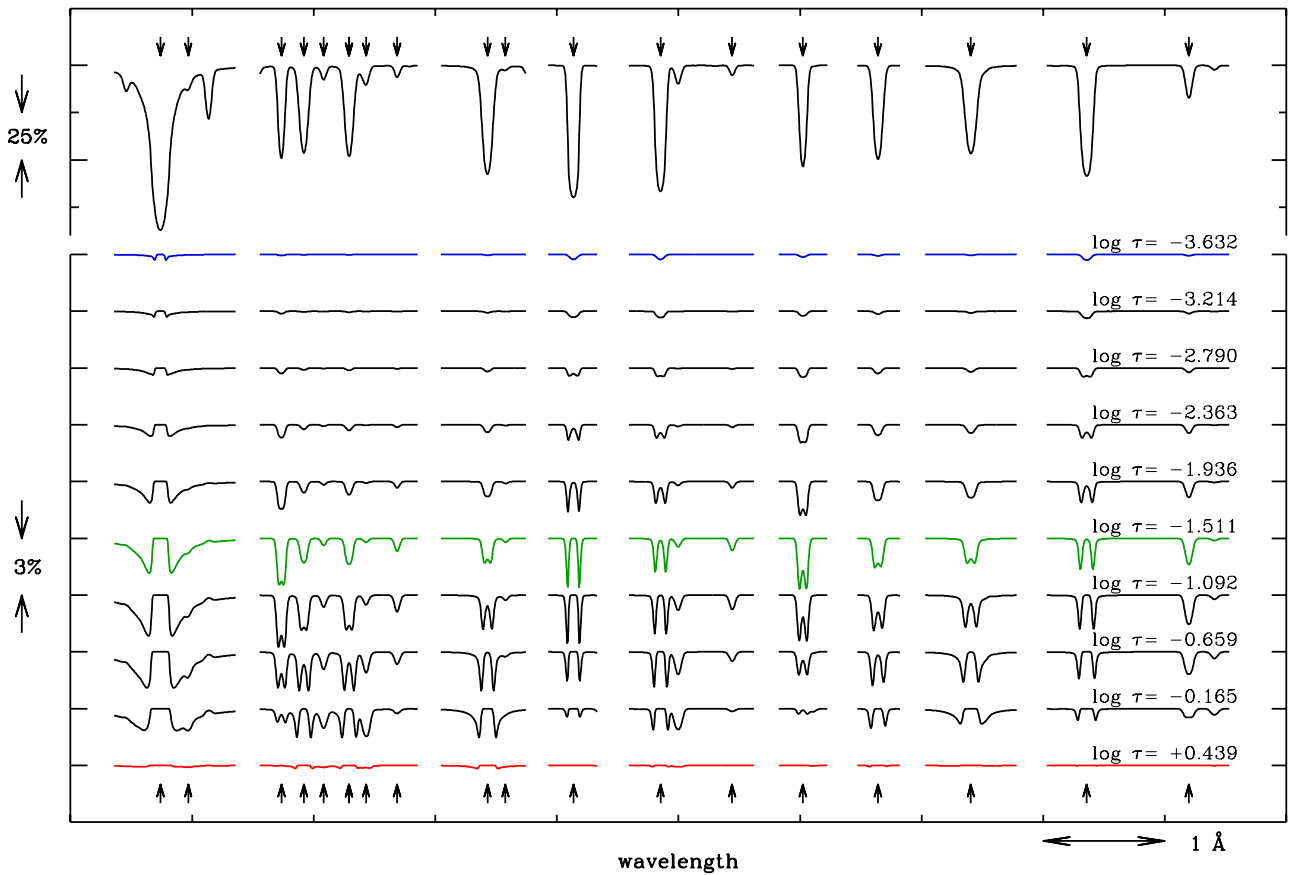}
\caption{{\bf Lower part:} The response of 19 Fe lines (indicated 
by arrows and listed in Table\,1) to perturbations of the abundance 
profile of KTR06 at a given optical depth. The effect of the 0.3\,dex 
perturbation -- see text and Fig.\,1 for details -- is given in the 
sense perturbed minus unperturbed Stokes $I$. {\bf Upper part:} The
unperturbed spectrum with the Fe lines.}
\label{fig2}
\end{figure*}

We do not want to deny that many of the stratification profiles presented 
in recent times lead to improved fits to the observed Stokes\,$I$ spectra, 
compared to an assumed constant abundance with depth. But it is also a fact 
that none of the predicted spectra are perfect, that residuals of 1-5\% 
in normalised intensity -- sometimes almost 10\% -- persist and that we 
stumble over lines where a stratified abundance gives a less satisfactory 
fit than a constant abundance. In the particularly well-studied star 
HD\,133792, for example, 3 out of 7 strontium lines belong to this category, 
and 5 out of 26 chromium lines. In the same star, the stratification of 
calcium has essentially been derived from just 3 lines, one of which is 
still rather poorly fitted by the stratification profile. Is it possible
to guarantee the uniqueness of a solution when the 
uncertainties in the atmospheric parameters, the limited accuracy of the 
atomic data, and in some cases also the unknown magnetic geometry of the 
star are kept in mind? Are the residuals due to the imperfect data or rather 
to the imperfect (and even possibly erroneous) stratification profiles?

In order to assess the question of the uniqueness of empirical stratification 
profiles it is imperative to first carry out a meaningful error analysis. 
Since HD\,133792 is the star for which the most detailed, and in a certain 
sense, the most sophisticated, determination of stratification profiles has 
ever been attempted, the VIP based stratification profiles of this star are 
certainly well suited for this purpose.

\subsection{Error analysis of the VIP}
\label{seq:VIP}

The stratification profiles for HD\,133792 have been determined by KTR06 under 
the assumptions of constancy and maximum smoothness. Their observations extend
over 10 minutes and cover just 1 phase. The quality of the fit to the observed 
spectrum varies between the elements. Differences between observed 
and predicted spectra reach some 2\% in about 10 of the 28 iron lines analysed
(we made these estimates from the figures). The centres of weak lines are 
affected to the same extent as the centres of much stronger lines; we also note 
that not all fits to line wings are fully satisfactory. The other elements do
not fare quite so well; whereas residuals can attain 5\% for Mg and Ca and 
6\% for Si 6\%, they exceed 9\% for Sr.

\begin{table}
\begin{tabular}{|rcc|rcc|rc|}
\hline
 ion & $\lambda$ & $\mid$ & ion & $\lambda$ & $\mid$ & ion & $\lambda$  \\
\hline
Fe 2 & 5018.440 & $\mid$ & Fe 1 & 5022.931 & $\mid$ & Fe 1 & 5434.524 \\
Fe 2 & 5018.669 & $\mid$ & Fe 1 & 5023.186 & $\mid$ & Fe 2 & 5567.842 \\
Fe 1 & 5022.236 & $\mid$ & Fe 2 & 5030.630 & $\mid$ & Fe 2 & 5961.705 \\
Fe 1 & 5022.420 & $\mid$ & Fe 2 & 5030.778 & $\mid$ & Fe 2 & 6149.258 \\
Fe 1 & 5022.583 & $\mid$ & Fe 1 & 5269.537 & $\mid$ & Fe 2 & 6150.098 \\
Fe 1 & 5022.789 & $\mid$ & Fe 2 & 5325.553 & $\mid$ &     &          \\
Fe 1 & 5022.792 & $\mid$ & Fe 1 & 5326.142 & $\mid$ &     &          \\
\end{tabular}
\caption{The iron lines shown in Fig.\,\ref{fig2}, listed with increasing
wavelength and indicated (from left to right) by arrows in the figure.}
\end{table}

At this point we shall not question the atmospheric model which could conceivably 
be of different effective temperature and gravity (Cowley, private communication), 
we neither question nor even take into consideration the magnetic field strength 
and geometry, we just take the published atmospheric parameters and the 
stratification profiles at face value. Sr is omitted in our investigation because 
of the exceptionally large residuals, but for the remaining elements we first
calculate the reference line spectrum predicted from the published stratification 
profiles. Then, in a controlled experiment, we determine just how large 
perturbations to these abundance profiles would have to be to lead to 1\% or 5\% 
deviations from the reference line spectrum. This in turn allows us to judge the 
significance of any detailed structure in the stratification profiles and makes 
it possible to estimate the interval over which these profiles are well defined. 
It is not surprising -- in view of the assumptions underlying the VIP approach -- 
that there is little such structure and that the remarkable smoothness of the 
empirical stratifications of Mg, Si, Ca, Cr and Fe (see Fig.\,5 of KTR06) is 
only slightly perturbed by humps or dips. The occurrence of gradients in the 
stratification profiles for $\log \tau_{\rm 5000} > 1$ however constitutes a major 
puzzle, given the basic physics of radiative transfer (see section 
\ref{seq:artefact} for a detailed discussion).

\begin{figure*}
\includegraphics[width=176mm]{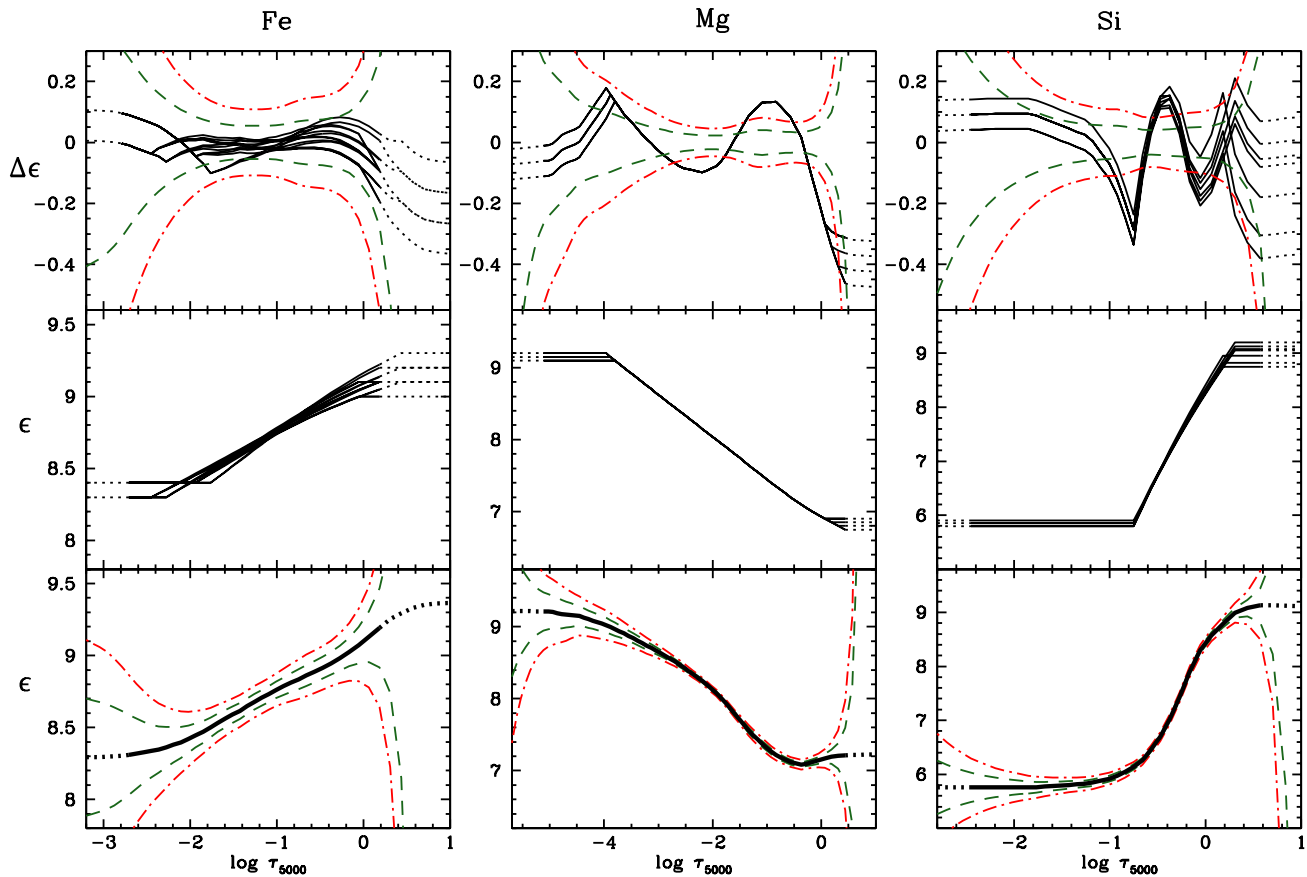}
\caption{{\bf Bottom panels:} Estimated uncertainties in the stratification 
curves of Fe, Mg, and Si. The dashed curves correspond to a $\pm 0.5$\% maximum
response to the perturbation in the spectrum, the dot-dashed curves to a 
$\pm 1$\% maximum response. The full lines are the respective original curves 
digitised from KTR06; they are plotted with dots where the 1\% confidence 
intervals become larger than 0.5\,dex. Minor wiggles as in the case of Fe are 
due to imperfect digitisation.
{\bf Middle panels:} Possible alternative stratifications for which the 
synthetic spectra do not differ by more than 0.002 - 0.003 (rms) and by 
less than 0.01 (maximum) from the spectra synthesised with the respective 
original stratification curves.
{\bf Top panels:} The same alternative stratifications plotted in the sense
alternative minus original profile. The respective dashed and dot-dashed
curves correspond to the confidence intervals displayed in the bottom panels.}
\label{fig3}
\end{figure*}

\subsubsection{The tools}
\label{seq:tools}

We chose a straightforward approach consisting in the application of 
a simple perturbation to the published stratification profiles. The 
Atlas12 code (Kurucz 2005) was used to establish an atmospheric model
for HD\,133792 with $T_{\rm eff} = 9400$\,K, $\log g = 3.80$ and solar 
metal abundances increased by 0.5\,dex. This does not constitute 
a perfect match to the model established by KTR06 but 
is largely sufficient for our purposes. This model atmosphere -- with 
99 layers -- was then used in the COSSAM polarised spectral synthesis
code (Stift 1998, 2000; Wade et al. 2001) to establish theoretical
spectra covering all the lines used by KTR06. The
atomic line data were taken from the VALD database (Piskunov et al. 
1995; Kupka et al. 1999). The public version of COSSAM provides solely 
a spatial integration grid centred on the line of sight and covering 
the visible hemisphere of the star, but for the present calculations 
we employed a corotating grid (Stift 1996) largely identical to those 
in general use in Doppler mapping (see Voigt, Penrod \& Hatzes 1987 for 
details). Taking the said 99 layer Atlas12 model for HD\,133792, a 5-layer 
perturbation of \{0.1 0.2 0.3 0.2 0.1\}\,dex was added to the empirical 
Fe profile of KTR06 (see Fig.\,\ref{fig1}) and a 
\{0.2 0.4 0.6 0.4 0.2\}\,dex perturbation to the stratification profiles 
of the other elements. Applying these perturbations in turn to all depth 
points, we determined the difference between original and perturbed spectrum. 
Fig.\,\ref{fig2} displays these differences for a selection of 19 Fe 
lines and for 10 different points in optical depth. For the 0.3\,dex 
perturbation to the iron stratification -- which corresponds to 1/3 of 
the total amplitude claimed by KTR06 and which is illustrated in 
Fig.\,\ref{fig1} -- the maximum effect on the normalised spectrum 
is of the order of 3\%. As expected, in the deeper layers the wings of 
strong lines provide most of the abundance information whereas in the 
upper layers this is mostly done by the line cores. Outside the interval 
$-2.5 < \log \tau_{\rm 5000} < -0.20$ the spectral response does not even 
reach 1\%, dropping rapidly below 0.5\% for $-2.9 < \log \tau_{\rm 5000}$
and for $\log \tau_{\rm 5000} > 0.0$. In other words, any attempt to 
reconstruct the Fe stratification profile beyond these limits cannot 
possibly yield meaningful results, simply because for the particular 
atmosphere in question and the abundance profile on which our 
calculations are based, the selected Fe lines become insensitive to 
abundance changes.

\subsubsection{Confidence intervals}
\label{seq:tools}

The lower panels of Figs.\,\ref{fig3}a-c show ``confidence intervals'' 
-- as derived with the above mentioned perturbative approach -- for the
stratification curves of Fe, Mg, and Si. These confidence intervals 
must not be confounded with those derived from rigorous statistics but 
result from the simple inversion of the relation perturbation vs. 
maximum response obtained from our calculations. A 0.5\% maximum response 
of the normalised spectrum requires a perturbation whose size is given 
by the distance between the dashed lines and the original profile. The 
perturbation necessary for a 1\% maximum response is reflected by the 
dot-dashed lines. It transpires from these results that the stratification 
profiles are well defined over relatively narrow intervals in optical depth 
-- here they are good to 0.1\,dex in the Fe case, and good to 0.2\,dex in 
the other cases -- but become essentially undefined for 
$\log \tau_{\rm 5000} > +0.6$. The narrowest such intervals with less than 
3~decades in optical depth are found for Fe and for Si, the Mg profile 
is defined over about 4~decades. 

We very carefully checked that the vertical 
resolution of the atmospheric model is good enough so that the response 
curves are not affected by numerical problems. For that purpose, the same 
analysis was carried out with a 199 layer Atlas12 model and a 9-layer 
perturbation, yielding essentially identical results.

\subsubsection{Assumptions and artefacts}
\label{seq:artefact}

We have seen from Figs.\,\ref{fig3}a-c that the portions of the abundance 
profiles below $\log \tau_{\rm 5000} = +0.6$ are invariably esentially undefined.
Still, gradients in the plotted abundance profiles for Fe, Si, Sr and Ca 
are clearly visible beyond $\log \tau_{\rm 5000} = 2.0$ in Figs.\,3a and 5 
of KTR06 and the question arises as to whether these have 
to be considered real or whether they rather constitute artefacts. From 
the formal solution of the radiative transfer equation the answer is 
unequivocally in favour of the latter: the contribution to the emerging 
intensity of the source function in the deepest layers multiplied by a 
factor of $\exp (-200)$ is {\em tens of decades smaller} than the 
contribution from the region near $\log \tau_{\rm 5000} = 0.0$.

In the classical step-function fitting procedure a transition region 
connects the respective lower and upper parts of the atmosphere which 
exhibit different constant abundances. Self-consistent diffusion models
(LeBlanc \& Monin 2004) display a similar structure. The VIP method 
instead starts with a constant abundance throughout the star and attempts 
to derive {\em deviations} from this mean abundance. Such profiles 
-- which converge to the same abundance value deep in the atmosphere 
and in the outermost layers, and which simply deviate from this value in 
some intermediate zone -- are neither in accord with the theoretical 
models of LeBlanc \& Monin (2004) nor with equilibrium solutions in the 
presence of magnetic fields presented by Alecian \& Stift (2007, 2008). 
The mentioned theoretical results are also at variance with the extreme 
smoothness of the stratification profiles. Disturbingly, Fig.\,3a of 
KTR06 shows that asymptotically their solution joins the mean abundance. 
A large regularisation parameter can make this asymptotic behaviour less 
visible, but it still persists and is readily visible in Fig.\,5 of KTR06. 
Choosing the regularisation parameter such that one arrives at more or 
less the same solution in the interval $-4.0 < \log \tau_{\rm 5000} < 0.5$ 
irrespective of the initial abundance guess (called 'optimum regularisation' 
by KTR06) constitutes a constraint that stabilises the solution but is
not based on any physical considerations. Such a procedure surely smoothes 
out spurious structure in the profiles and effectively hides the undesirable 
asymptotic behaviour in the deepest layers, but there is no way to assess 
to what degree this might result in unwarranted and potentially severe 
smearing out of the abundance jump. 

It should also be mentioned that a basic assumption underlying all 
empirical approaches so far towards the derivation of abundance 
stratifications in Ap stars, i.e. the insensitiveness of diffusion to 
magnetic fields, appears to be in serious contradiction with established 
theoretical wisdom. It has been shown already by Alecian \& Vauclair (1981) 
that strong horizontal magnetic fields do have an impact on diffusion in 
stellar atmospheres. Even in a field of only 1\,kG, stratification profiles 
start to depend on field direction for $\log \tau_{\rm 5000} < -3$, see 
Fig.\,3 of Alecian \& Stift (2007). So nobody can reasonably exclude that 
despite the rather moderate field of HD\,133792, the outer parts of the 
stratification profiles of Sr, Ca, and Mg may be affected. And 
certainly one should not overlook the studies by Kurtz, Elkin \& Mathys 
(2005) which suggest a concentration of rare earth elements at 
$\log \tau_{\rm 5000} = -4$ and higher in the atmosphere.
In general, as 
Alecian \& Stift (2008) have shown, equilibrium stratifications are quite 
sensitive to strong ($> 5$\,kG) magnetic fields as found in $\beta$\,CrB, 
$\gamma$\,Equ, HD\,144897, and HD\,66318 to mention just a few of the 
well-studied stars. In this context it should always be kept in mind that it 
is not the 2-fold difference in field strength between pole and equator in 
a dipolar oblique rotator model that is decisive, but the direction of the 
field. Results based on Zeeman Doppler mapping which suggest that abundance
anomalies are related to the magnetic field topology (see e.g. Kochukhov 
et al. 2002) are certainly in qualitative accord with theoretical findings.

\section{Uniqueness of empirical stratifications}
\label{seq:unique}

The claim that a model derived with the ``optimum regularisation'' is 
unique has to be understood in the sense that it is unique within the 
framework of VIP (Kochukhov, private communication). These seems to
be a reasonable claim, but we really want to have a look at the 
uniqueness of empirical stratification profiles in a more general 
sense. Looking at HD\,133792, are there other abundance profiles that 
reproduce the observed spectrum as well as the VIP solution? Faced 
with an infinity of possible profile shapes and profile distributions,
we started with the simplest case, viz. a step function whose shape 
stays constant over the star, irrespective of magnetic latitude.

\begin{figure*}
\includegraphics[width=176mm]{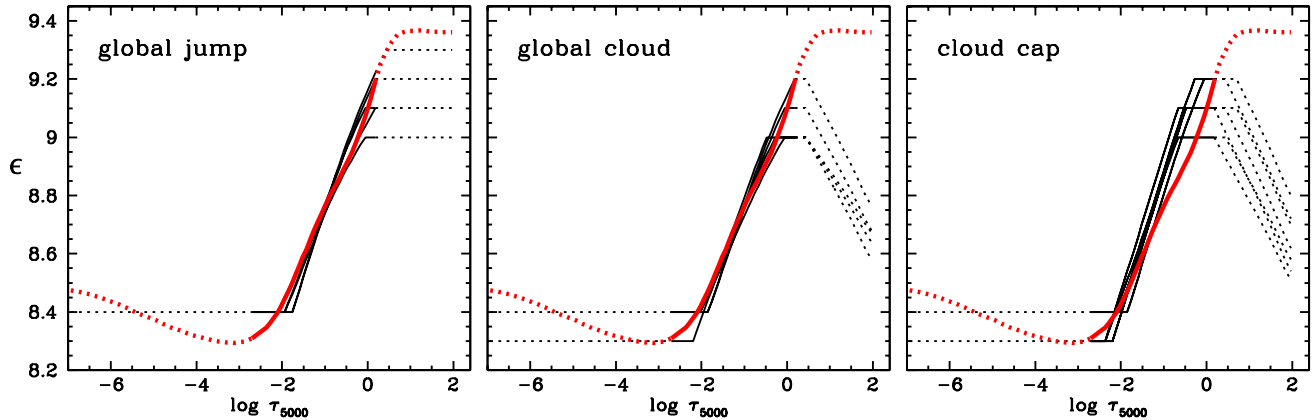}
\caption{{\bf (a)} The smooth original digitised VIP stratification curve 
for Fe together with a selection of global polygonal curve step-function 
alternatives. The curves are plotted with dots where the confidence 
interval exceeds 0.5\,dex (as explained in Fig.\,\ref{fig3}).
{\bf (b)} The same for global cloud models and {\bf (c)} for
clouds confined to a polar cap.}
\label{fig4}
\end{figure*}

A 4 parameter step function is defined by the ``outer'' and the ``inner'' 
abundances, and by the position and width of the transition region
which connects the ``inner'' and the ``outer'' parts.
We made a reasonably extended, almost exhaustive search for such step
functions in the case of Mg, Si, Ca, and Fe, by calculating a dense 
grid covering all possible parameter values. Step function solutions
were considered acceptable whenever the rms deviation of the alternative 
spectrum from the normalised spectrum calculated with the VIP solution 
did not exceed $0.002 - 0.003$ and when the maximum deviation was less 
than 1\%. Given the 2 - 6\% maximum size of the residuals of the VIP fits 
for the elements in question, this is indeed an extremely strong constraint. 
The middle panels of Figs.\,\ref{fig3}a-c display a selection of acceptable 
global step function solutions. For clarity they are plotted again in the 
top panels in the sense alternative minus VIP solution, and in addition, 
confidence intervals overlay the curves. At this point we want to 
remind the reader again that we are in no way looking for alternative 
fits to the {\em real observed spectrum}, but only for excellent fits to 
the {\em VIP based synthetic spectrum}. So one would not expect the 
alternative stratifications to differ completely from the VIP profiles.

In the case of Fe, the differences between step function solutions and the
VIP curve tend to stay within the 0.5\% response curve. Still, for the 
large number of acceptable step function models we find a noticeable spread 
in amplitude and also in the width of the transition region, due to the 
virtual lack of response to perturbations for $\log \tau_{\rm 5000} > 0.2$. 
It thus becomes impossible to determine by line-profile fitting where 
deep in the atmosphere the transition region for Fe ends. But, as so often 
in the diagnosis of stellar atmospheres, seismology might possibly offer 
some hope for the future.
Mg and Si do not exhibit quite the same simple behaviour, and differences 
between the respective alternative step function models and the original 
curves extend consistently and sometimes substantially beyond the 1\% 
curves. Compared to the mean amplitude and to the mean transition 
region width of the alternative profiles for Mg, the spread in these
quantities is relatively small; this spread becomes slightly more 
important for Si.

These findings confirm the correctness of our conjecture that not even for 
a spectrum fit at the 0.5\% accuracy level can a unique model be guaranteed: 
neither does a 0.3\,dex spread in the Fe abundances near the bottom of the 
atmosphere influence the fit to the spectrum, nor does a 0.6\,dex 
amplitude give less satisfactory results than the 1.05\,dex VIP amplitude.

\subsection{Clouds, caps and rings}
\label{seq:cloud}

Many more simple abundance profiles can be explored which constitute 
a zero-order approximation to equilibrium stratification profiles 
found for magnetic stellar atmospheres. We decided to have a look at 
one particular family of models which are based on the assumption that 
a cloud of increased elemental abundance hovers somewhere in the 
atmosphere. This cloud is either distributed uniformly all over the 
star, or restricted to a cap around one magnetic pole, or located in 
a ring around the magnetic equator. There are 2 transition regions,
one towards the upper part of the atmosphere, the second towards
the bottom, outside of which the abundance is constant and not 
necessarily solar. As it turns out, all these geometries can lead to 
excellent fits to the VIP based synthetic spectrum. Figs.\,\ref{fig4}a-c 
compare the VIP iron profile to global step-function (jump) solutions as 
discussed above, to global cloud models, and to clouds confined to a 
polar cap. In all 3 classes of models shown, the abundances in the 
deeper layers are not particularly well defined, in contrast to the 
abundances in the outer layers. In the global step-function case 
(Fig.\,\ref{fig4}a) the 0.3\,dex spread in abundances at the bottom 
of the atmosphere corresponds to a remarkable 50\% of the minimum 
possible value of the jump; in the cases of a global cloud 
(Fig.\,\ref{fig4}b) and of a cloud in a polar cap (Fig.\,\ref{fig4}c) 
the spread at the bottom reduces to 0.2\,dex -- the minimum amplitude 
remaining at 0.6\,dex. For the ring case we only dispose of a few 
hundred models and therefore it is not possible to give definitive 
values for the total possible spread in amplitude.

We did not delve into an exhaustive search for cap models at all
possible phases but concentrated on cap models seen pole-on and on 
ring models seen equator-on. In both cases we found a rather narrow 
range of extensions, caps covering the star up to $50 - 60\degr$ 
from one magnetic pole, and rings being confined to $25 - 35\degr$ 
from the magnetic equator. Cap solutions consistently give narrower 
transition regions than both global step-function and global cloud 
solutions; for a given optical depth, the abundance is substantially 
higher in the interval $-1.5 <  \log \tau_{\rm 5000} < 0$ (compare
Figs.\,\ref{fig4}ab to Fig.\,\ref{fig4}c). This behaviour seems to be 
followed by the existing ring solutions.

Comparison of the VIP amplitude of about 1.05\,dex with the maximum 
possible amplitude of 0.9\,dex (global step-function case) or 0.8\,dex 
(cloud cases) and with the minimum possible amplitude of 0.6\,dex
-- in conjunction with the remarkable spread in abundance at the bottom 
of the atmosphere -- reveals just how uncomfortably large the 
uncertainties in the empirical profiles really are.

\section{Conclusions}
\label{seq:clus}

On a positive note, our modelling confirms that it is possible to
unequivocally establish the presence of chemical stratifications and that 
the sense of the abundance change is always clear. There can be thus no 
doubt that the decrease with depth of the Mg abundance in HD\,133792, for 
example, is indeed a decrease, and that the abundance of Fe increases with 
depth in this star. The respective {\em orders of magnitude} of the 
abundance jumps are quite well defined.

Our findings however reveal that none of the presently used approaches
is capable of providing unique stratification profiles. Taking the 
slowly rotating Ap star HD\,133792, we have shown that for several 
chemical elements, many different global step-function-like solutions 
can be found which perfectly reproduce the VIP based synthetic 
spectrum; the respective amplitudes of the jumps however can differ 
substantially among each other. For Fe the step-function amplitudes 
are invariably smaller than the VIP value. We have further shown that 
the quality of the VIP fit is also well-matched by either global 
cloud-like solutions, by clouds in a cap around a magnetic pole or by 
clouds in a ring about the magnetic equator. Both for cap and ring 
models, we find a certain spread in amplitudes and in addition a 
narrowing of the transition region. Even when, as in the case of 
HD\,133792, profiles of 28 Fe lines are used in the inversion, there 
appears to be no way to distinguish between the various possible 
stratification profiles (at least not with Stokes $I$ only).

Cap and ring geometries can be seen as rough approximations to the 
results of equilibrium stratification calculations by Alecian \& Stift 
(2008) who have demonstrated that in magnetic fields of 5\,kG and more, 
stratification profiles become strongly dependent on the field angle.
While we consider that in strongly magnetic stars like HD\,66318 and 
HD\,144897 these models could possibly come slightly closer to reality 
than models which assume globally constant profile shapes, nobody can
guarantee uniqueness or even correctness. Only with excellent phase
coverage instead of observations at just 1 phase, and with high quality
observations in all 4 Stokes parameters will it perhaps become possible 
to distinguish between the rival models. Ideally one would have to
reconstruct the run of abundance with depth and the magnetic field 
vector at each point of the stellar surface. Whether such an extremely 
ill-defined problem can ever be solved is hard to predict.

Thus, at present, surveys of (magnetic) Ap stars in view of empirically 
establishing the {\em extent of the stratification phenomenon} are invaluable 
for our understanding of radiatively driven diffusion and its dependence 
on stellar parameters including magnetic fields. Empirical inversions
reveal the sense and the order of magnitude of an abundance change with 
depth, but not the exact amplitude, nor the precise location of the 
transition region, and certainly not any fine structure in the stratification 
profile. Usually empirical stratifications are only defined over a rather
restricted interval in optical depth, apparently never beyond 
$\log \tau_{\rm 5000} > +0.6$, and for the reasons discussed above, they
are expected to be particularly unreliable in strongly magnetic Ap stars. 
They cannot therefore in the foreseeable future provide the desired strong 
constraints to theoretical diffusion modelling.

\section*{Acknowledgements} 
MJS acknowledges support by the {\sf\em Austrian Science Fund (FWF)}, 
project P16003-N05 ``Radiation driven diffusion in magnetic stellar 
atmospheres''. Dr. Shulyak generously provided his Linux version of 
the Atlas12 code and kindly helped with the installation. Thanks also 
go to Dr. Kochukhov for most interesting discussions and a number of 
clarifications. Helpful comments by the referee have improved 
the manuscript.

\bsp
	
{}

\label{lastpage}

\end{document}